\documentclass[runningheads]{llncs}
\usepackage{graphicx}
\usepackage{multirow}
\usepackage{array}
\usepackage{amsmath,amssymb,amsfonts}
\usepackage[perpage]{footmisc}
\usepackage[misc]{ifsym} 

\begin{document}
%
% \title{Contribution Title\thanks{Supported by organization x.}}
\title{PEN4Rec: Preference Evolution Networks for Session-based Recommendation}

% \author{Anonymous author(s)}
%
\titlerunning{Preference Evolution Networks for Session-based Recommendation}
% todo: anonymous submission
\author{Dou Hu\inst{1} 
\and Lingwei Wei\inst{2,3}
\and Wei Zhou\inst{2}(\Letter)
\and Xiaoyong Huai\inst{1} 
\and Zhiqi Fang\inst{1} 
\and Songlin Hu\inst{2,3}
}
\authorrunning{Dou Hu et al.}

\institute{
National Computer System Engineering Research Institute of China 
\and
Institute of Information Engineering, Chinese Academy of Sciences
\and 
School of Cyber Security, University of Chinese Academy of Sciences \\
% \email{ \{hudou18,weilingwei18\}@mails.ucas.edu.cn } \email{\{zhouwei,husonglin\}@iie.ac.cn}
% \email{ huaixy@sina.com }
\email{zhouwei@iie.ac.cn}
}

\maketitle              % typeset the header of the contribution
\begin{abstract}
  \let\thefootnote\relax\footnotetext{\Letter { } Corresponding author.}

Session-based recommendation aims to predict user the next action based on historical behaviors in an anonymous session. For better recommendations, it is vital to capture user preferences as well as their dynamics. Besides, user preferences evolve over time dynamically and each preference has its own evolving track. However, most previous works neglect the evolving trend of preferences and can be easily disturbed by the effect of preference drifting. In this paper, we propose a novel Preference Evolution Networks for session-based Recommendation (PEN4Rec) to model preference evolving process by a two-stage retrieval from historical contexts. Specifically, the first-stage process integrates relevant behaviors according to recent items. Then, the second-stage process models the preference evolving trajectory over time dynamically and infer rich preferences. The process can strengthen the effect of relevant sequential behaviors during the preference evolution and weaken the disturbance from preference drifting. Extensive experiments on three public datasets demonstrate the effectiveness and superiority of the proposed model.

\keywords{ Recommender systems
\and Session-based recommendation 
\and Graph neural networks 
\and Sequential behavior
\and User preference.}
\end{abstract}
\section{Introduction}
With the rapid development of online platforms, \textit{Recommender Systems} \cite{DBLP:conf/www/LiuRZC0Y20,DBLP:journals/ml/ZhangLG19} have received increasing concern recently.
In many real-world services, user identification is not always available and only the historical user-item interactions during an ongoing session can be accessed easily.
In such scenario, \textit{Session-based Recommendation} (SBRS) \cite{DBLP:conf/sigecom/SchaferKR99} emerges by focusing on an anonymous session of user-item interactions (e.g., purchases of items) within a certain period. 
Exploring user preferences as well as their dynamics behind user-item interactive behaviors in the session is the key to advance the performance of SBRS.

Traditionally, similarity-based \cite{DBLP:conf/www/SarwarKKR01} and matrix factorization \cite{DBLP:conf/uai/RendleFGS09,DBLP:reference/rsh/KorenB11} methods are not suitable for the session-based scene because of ignoring the order of the user's behaviors.
Some methods \cite{DBLP:conf/www/RendleFS10,DBLP:journals/corr/HidasiKBT15} deal with dependencies between adjacent behaviors successively and equally. 
Intuitively, not all behaviors are strictly dependent on each adjacent behavior. 
Each user has diverse preferences, and each preference has its own evolution.
That is, the user's intentions can be very different in adjacent visits, and one behavior of a user may depend on the behavior that takes a long time ago. Such a phenomenon can be named \textit{preference drifting},  caused by the diversity and dynamics of user preferences.
Recently, many works \cite{DBLP:conf/cikm/LiRCRLM17,DBLP:conf/sigir/WangRMCMR19,DBLP:conf/kdd/LiuZMZ18,DBLP:conf/aaai/WuT0WXT19,DBLP:conf/ijcai/XuZLSXZFZ19} apply an attention mechanism to integrate relative behaviors. Although they can capture diverse preferences, they still ignore sequential patterns in different preferences and obtain one fixed preference evolving trajectory. Thereby, they can be disturbed by the preference drifting.

In  this paper, we propose a novel two-stage {\bf P}reference {\bf E}volution {\bf N}etworks {\bf for} session-based {\bf Rec}ommendation (short for {\bf PEN4Rec}), to model preference evolving process based on historical contexts.
Following graph-based models \cite{DBLP:conf/aaai/WuT0WXT19,DBLP:conf/ijcai/XuZLSXZFZ19}, we first encode user-item behaviors in a session graph to capture complex item transitions under multi-hop neighbor connections. 
After that, different from taking behaviors as preferences directly, we extract user preferences via a well-designed two-stage retrieval. 

Specifically, the {first-stage} process integrates relevant behaviors from historical contexts according to the recent items via the attention mechanism.
And the {second-stage} process models the preference evolving trajectory over time dynamically and reasons diverse preferences.
There are two key layers in the second-stage process, i.e., a session reader layer and a preference fusion layer.
The former layer applies an adaptive bidirectional gated recurrent unit (Bi-GRU) to sufficiently gather the contextual information for each item in two directions of the ongoing sequential session.
Then, the latter layer combines the attention mechanism and gated recurrent unit (GRU) to capture more relevant preferences during the preference evolution. 
The attention weights are applied to strengthen the above relevant behaviors' influence on the preference evolution and weaken irrelevant behaviors' effect that results from preference drifting. 
Through the above two-stage retrieval, we can produce more diverse preferences and make precise recommendations.

To assess the proposed model, we compare with typical baselines on three public benchmark datasets. Experimental results demonstrate the effectiveness of the proposed model. 
% Overall, 
The main contributions are summarized as follows:
\begin{itemize} 
  \item We investigate the preference drifting phenomenon for session-based recommendations, and propose a new two-stage PEN4Rec to model preference evolving process based on historical contexts. 
  \item We design a novel second-stage retrieval process to strengthen the effect of relevant sequential behaviors during the preference evolution and weaken the disturbance from preference drifting.
  \item We conduct extensive experiments on three real-world benchmark datasets.
  The results consistently demonstrate the effectiveness and superiority of the proposed model\footnote{The source code is available at \url{https://github.com/zerohd4869/PEN4Rec}}.
\end{itemize}

% The rest of this paper is organized as follows. Section \ref{sec:related} presents a brief review of the related work. Section \ref{sec:task} formulate the task of session-based recommendation. In Section \ref{sec:model}, we describe the proposed model in detail. Then, we report experimental settings in Section \ref{sec:experiment} and analyze the results in Section \ref{sec:results}. Section \ref{sec:conclu} concludes the paper and points out future research directions

\section{Related Work} \label{sec:related}

\subsection{Conventional Recommendation Methods}
Similarity-based methods \cite{DBLP:conf/www/SarwarKKR01}  recommended items similar to the previously clicked item in the session.
Matrix factorization methods \cite{DBLP:conf/uai/RendleFGS09,DBLP:reference/rsh/KorenB11} represented user preferences by factorizing a user-item matrix consisting of the whole historical behaviors.
These methods are not suitable for the session-based scene because of ignoring the order of the user's behaviors.
Then, Rendle et al. \cite{DBLP:conf/www/RendleFS10} combined Markov chain and matrix factorization to simulate the sequential behavior between two adjacent interactions while ignoring long-term dependencies.

\subsection{Deep Learning based Recommendation Methods}
In recent years, many deep learning based methods are proposed for SBRS.
Particularly, 
Hidasi et al. \cite{DBLP:journals/corr/HidasiKBT15} first employed recurrent neural networks (RNNs) \cite{1986Serial} to simply treat the data as time series.
The work had facilitated the investigation of RNN-based models \cite{DBLP:conf/aaai/RenCLR0R19} in SBRS.
Li et al. \cite{DBLP:conf/cikm/LiRCRLM17} applied an attention mechanism on RNN to capture sequential features and main intents. % NARM
Liu et al. \cite{DBLP:conf/kdd/LiuZMZ18} used an attentive network to capture short-term and long-term preferences.
Wang et al. \cite{DBLP:conf/sigir/WangRMCMR19} exploited the key-value memory networks \cite{DBLP:conf/emnlp/MillerFDKBW16} to consider information from the current session and neighborhood sessions.
More recently, some graph-based works \cite{DBLP:conf/aaai/WuT0WXT19,DBLP:conf/ijcai/XuZLSXZFZ19,DBLP:conf/kdd/ChenW20,DBLP:conf/cikm/PanCCCR20,DBLP:conf/cikm/QiuLHY19} apply graph neural networks (GNNs) \cite{DBLP:journals/corr/LiTBZ15,DBLP:journals/tnn/ScarselliGTHM09,DBLP:conf/iclr/VelickovicCCRLB18} to learn complex item transitions based on the session graph.
Especially, Wu et al. \cite{DBLP:conf/aaai/WuT0WXT19} and Xu et al. \cite{DBLP:conf/ijcai/XuZLSXZFZ19} extracted preferences after a gated graph neural network (GGNN) \cite{DBLP:journals/corr/LiTBZ15} by using the attention layer to capture long-term preferences.

Although these methods show promising performance, almost all the above approaches neglect the evolving trend of preferences and can be easily disturbed by the effect of preference drifting.
Different from them, the proposed PEN4Rec can strengthen the effect from relevant sequential behaviors during the preference evolution and weaken the disturbance effect that results from preference drifting.

\begin{figure*}[t]
  \centering
  \includegraphics[width=0.92\textwidth]{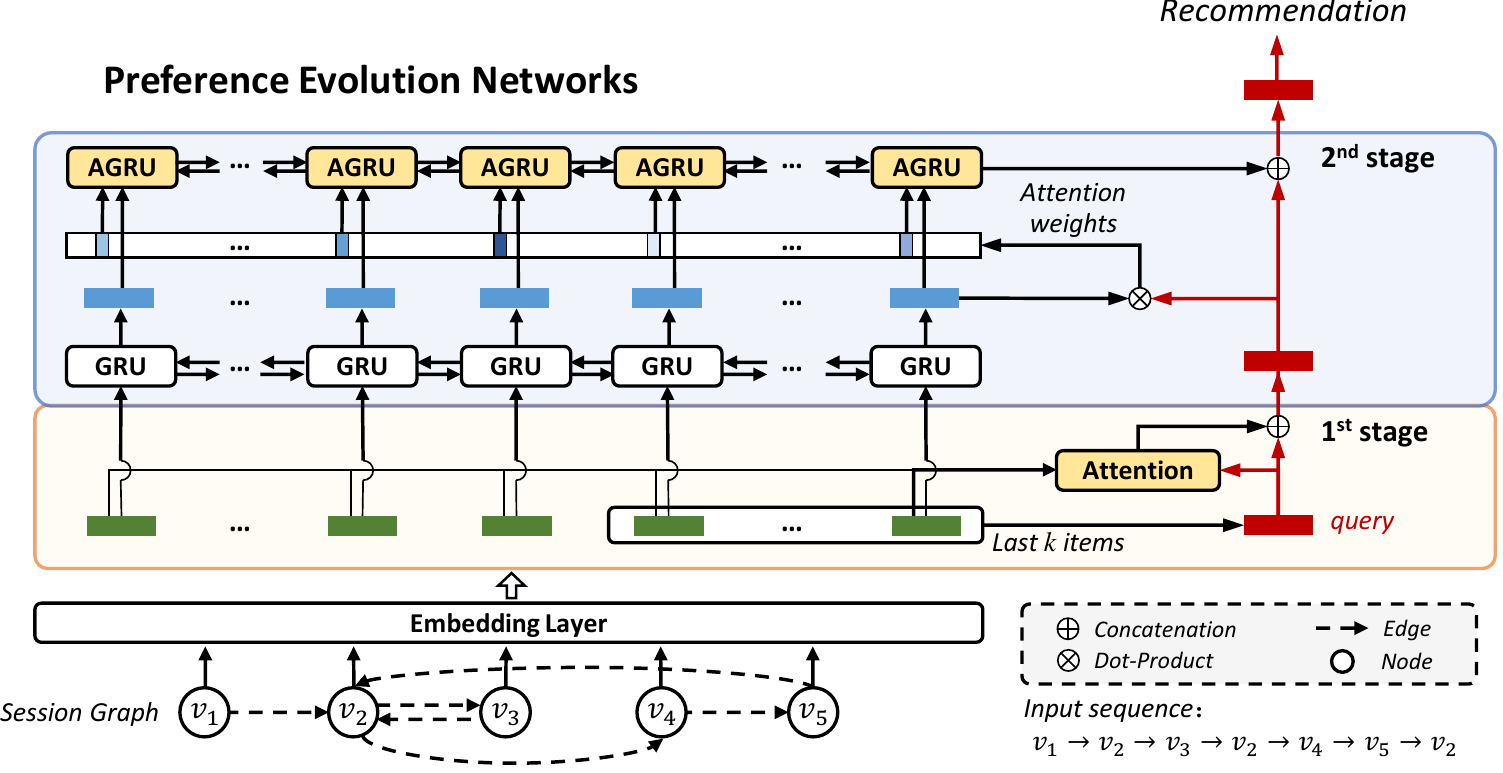}
  \caption{The architecture of the proposed model. PEN4Rec builds a directed session graph for each session sequence and encode items into a unified embedding space by the embedding layer. Then, in the first-stage process, irrelevant behaviors are integrated according to the last $k$ items from historical contexts. In the second-stage process, the model retrieves diverse preferences and reasons the preference evolving trajectory that is relevant to the updated query. Finally, PEN4Rec generates a vector representation with diverse preferences to make recommendations. 
  }
  \label{fig:overview}
\end{figure*}

\section{Problem Definition}  \label{sec:task}
In this section, We will define the task of session-based recommendation. 

Formally, let $\mathcal{V}=\{ v_1, v_2, ..., v_{|\mathcal{V}|} \}$ denote the set of all unique items involved in all sessions.
We define an anonymous session sequence $s$ as a sequence of items $s = [v_{1}, v_{2},...,v_{n}]$ ordered by timestamp, where $v_{i} \in \mathcal{V}$ represents a user-clicked item at time step $i$ within the session $s$. 
The goal of {\bf session-based recommendation} is to predict the next item, i.e., $v_{n+1}$, given the session $s$.
Generally, we design and train a model to output a probability distribution $p(v|s)$ over the entire item set $\mathcal{V}$.
The items with top-$c$ probabilities will be in candidate items for recommendation.

\section{The Proposed Model} \label{sec:model}
% . 
% The whole architecture of the proposed model is shown
In this section, we propose {\bf P}erference {\bf E}volution {\bf N}etworks for session-based {\bf Rec}ommendation (\textbf{PEN4Rec}), as shown in Fig.~\ref{fig:overview}.

\subsection{Session Graph Construction}
Firstly, given an input session sequence $s = [v_{1}, v_{2},...,v_{n}]$, we build a directed session graph $\mathcal{G} = (\mathcal{V}, \mathcal{E})$.
In the session graph, we treat each item $v_{i} \in \mathcal{V}$ as a node. 
Each edge $(v_{i-1}, v_{i}) \in \mathcal{E}$ is formulated as a directed edge to represent a user clicks item $v_{i-1}$ and $v_{i}$ consecutively. 
The connection matrix $\textbf{A} \in \mathbb{R}^{n \times 2n}$ is defined as the concatenation of two adjacency matrices, which are denoted weighted connections of outgoing and incoming edges in $\mathcal{G}$, respectively.  
Since several items may appear in the session sequence repeatedly, we assign each edge with normalized weighted. The value is computed as the occurrence of the edge divided by the out-degree of that edge's start.

\subsection{Embedding Layer}
Graph neural networks (GNNs) \cite{DBLP:journals/corr/LiTBZ15,DBLP:journals/tnn/ScarselliGTHM09,DBLP:conf/iclr/VelickovicCCRLB18} are well-suited for SBRS due to the powerful ability to feature extraction with considerations of rich node connections.
To weaken strong chronological order in the context of a session graph, we improve the gated graph neural network (GGNN) \cite{DBLP:journals/corr/LiTBZ15} with the attention mechanism to adaptively capture complex item transitions under multi-hop neighbors, which are difficult to be revealed by previous sequential methods \cite{DBLP:conf/www/RendleFS10,DBLP:journals/corr/HidasiKBT15,DBLP:conf/cikm/LiRCRLM17,DBLP:conf/sigir/WangRMCMR19,DBLP:conf/kdd/LiuZMZ18}.

Formally, each node $v_{i} \in \mathcal{V}$ can be embeded into an unified embedding space and the node vector $\textbf{v}_i \in \mathbb{R}^d$ indicates a $d$-dimensional latent vector of item $v_i$. 
We assume there are no strong sequential relations among $k$ successive items, e.g., $k$ can be the average number of continuously clicked items. Under this assumption, the $k+1$-th item may be related to a part of the previous $k$ items. The $t$-th layer of transitions between item nodes can be defined as:
\begin{equation}
% \resizebox{0.4\columnwidth}{!}{$
\begin{split}
    \textbf{a}^t_{c,i} &=  \textbf{A}_{c,i:} [\textbf{v}^{t-1}_1, ..., \textbf{v}^{t-1}_n]^\top \textbf{W}_{a}  + \textbf{b}_{a}, \\
  \alpha_c &= softmax ({({\textbf{W}_\alpha \textbf{a}^t_{c,i} })}^\top \textbf{v}^{t} ),   \\ 
  \textbf{a}^t_{i} &= \sum\nolimits_{c} {\alpha_c \textbf{a}^{t}_{c,i}},
\end{split}
% $}
\end{equation}
where $\textbf{A}_{c,i:} \in \mathbb{R}^{1 \times 2n}$ are the two columns of blocks in $\textbf{A}$ under $c$-hop neighbors corresponding to $v_{i}$, $c=k-1$.
% $\textbf{A}_{i:} \in \mathbb{R}^{1 \times 2n}$ are the two columns of blocks in $\textbf{A}$ corresponding to node $v_{i}$.
$\textbf{W}_{a} \in \mathbb{R}^{d \times 2d}$ and $ \textbf{b}_{a} \in \mathbb{R}^{d}$ are learnable parameters.
$\textbf{a}^t_{i}$ adaptively extracts the contextual information of multi-hop neighbors for $v_{i}$.
Then, the final output $\textbf{v}^{t}_i$ of the embedding layer is computed by:
 \begin{equation}
  \begin{split}
%     \textbf{a}^{t}_{i} &= 
%   \textbf{A}_{i:} [\textbf{v}^{t-1}_1, ..., \textbf{v}^{t-1}_n]^\top \textbf{W}_{a}  + \textbf{b}_{a}, \\
    \textbf{z}^t_{i} & = \sigma (\textbf{W}_z   \textbf{a}^t_{i} + \textbf{U}_z \textbf{v}^{t-1}_i), \\
  \textbf{r}^t_{i} & = \sigma (\textbf{W}_r   \textbf{a}^t_{i} + \textbf{U}_r \textbf{v}^{t-1}_i), \\
  \widetilde{\textbf{v}}^t_{i} & = tanh(\textbf{W}_o \textbf{a}^t_{i} + + \textbf{U}_o (\textbf{r}^t_{i}  \odot \textbf{v}^{t-1}_i )), \\
  \textbf{v}^{t}_i & = (1 - \textbf{z}^t_{i} ) \odot \textbf{v}^{t-1}_i + \textbf{z}^t_{i} \odot \widetilde{\textbf{v}}^t_{i}, 
  \end{split}
 \end{equation}
 where 
$\textbf{z}^t_{i}$ and $\textbf{r}^t_{i}$ are reset and update gates to determine the preserved and discarded information, respectively.
 $\textbf{W}_z, \textbf{W}_r, \textbf{W}_o \in \mathbb{R}^{2d \times d}$, 
 $\textbf{U}_z, \textbf{U}_r, \textbf{U}_o \in \mathbb{R}^{d \times d}$ 
 are learnable parameters.
 $\sigma(\cdot) $ is the sigmoid function.
 $\odot$ is the element-wise multiplication.

\subsection{Preference Evolution Networks}
Due to joint influence from the external environment and internal cognition, user preferences evolve over time dynamically.
PEN4Rec aims to capture user preferences and models preference evolving process by transitively retrieving relevant information in the two-stage process.
The advantages are two points. First, the model can supply the representation of recent preferences with more relative history information. Also, it is better to predict the next item by following the preference evolution trend.

\subsubsection{The First-stage Process} 
We design {the first-stage process} to integrate relevant behaviors from historical contexts according to the local preference. We regard recent $k$ items as the local preference since the next item may be related to a part of the previous $k$ items. Thus, the query is defined as $\textbf{q}^{(0)} = [\textbf{v}_{n-k+1};...;\textbf{v}_{n}]$.
Then, we aggregate relevant behaviors by adopting the soft-attention mechanism to better represent the global preference $\textbf{p}^{(0)}$:
\begin{equation}
  \begin{split}
  \beta_i &= \sigma (\textbf{W}_{1} \textbf{q}^{(0)}  + \textbf{W}_{2} \textbf{v}_i + \textbf{b} ), \\
  \textbf{p}^{(0)} &= \sum\nolimits_{i=n-k+1}^n \beta_i \textbf{v}_i,
\end{split} 
  \end{equation}
where $\textbf{W}_{1} \in \mathbb{R}^{d \times kd}$ and  $\textbf{W}_{2} \in \mathbb{R}^{d \times d}$ controls the weights of item embedding vectors, and $\textbf{b} \in \mathbb{R}^{d}$ is the bias parameter. 
After the first-stage process, the query can be updated as:
\begin{equation}
  \textbf{q}^{(1)} = \textbf{W}_q [\textbf{p}^{(0)}; \textbf{q}^{(0)}] + \textbf{b}_q,
\end{equation} 
where $\textbf{W}_q \in \mathbb{R}^{d \times 2d}$ and $\textbf{b}_q \in \mathbb{R}^d$ are parameters.

\subsubsection{The Second-stage Process}  
The second-stage process is designed to retrieve diverse preferences and reason the preference evolving trajectory that is relevant to the updated query via a {\it session reader layer} and a {\it preference fusion layer}. 

For {\bf the session reader layer}, we utilize an adaptive bidirectional GRU (Bi-GRU) to record the contextual information for each item in two directions.
Bi-GRU allows for message propagation from neighboring contexts, capturing spatial information in the ongoing sequential session.
Formally, given the item embedding  $\textbf{v}_i$, the retrieval can be formulated as:
\begin{equation}
  \begin{split}
  \overrightarrow{\textbf{m}}_{i} &= \overrightarrow{GRU}(\textbf{v}_i, \overrightarrow{\textbf{m}}_{i-1}), \\
  \overleftarrow{\textbf{m}}_{i} &= \overleftarrow{GRU}(\textbf{v}_i, \overleftarrow{\textbf{m}}_{i+1}),
\end{split}
\end{equation}
where $\overrightarrow{\textbf{m}}_{i}, \overleftarrow{\textbf{m}}_{i} \in \mathbb{R}^d$ indicates the hidden states of the forward GRU and the backward GRU. 
Then we adopt a residual connection by concatenating them and combine the input to enable sufficient interaction between items.
\begin{equation}
  \textbf{m}_{i} = tanh( \textbf{W}_m [\overrightarrow{\textbf{m}}_{i};\overleftarrow{\textbf{m}}_{i}] + \textbf{b}_m) + \textbf{v}_{i},
\end{equation}
where  $\textbf{W}_m \in \mathbb{R}^{d \times 2d}$ and $\textbf{b}_m \in \mathbb{R}^{d}$ are learnable parameters.

{\bf The preference fusion layer} retrieves the information to adaptively extract key information in complex queries, which is inspired by dynamic memory networks \cite{DBLP:conf/icml/XiongMS16}. 
It is formed by modifying the GRU by embedding information from the attention mechanism.
Specifically, we update the internal state of a normal GRU with the attention weights to strengthen the effect of relevant behaviors on the preference evolution and weaken irrelevant behaviors' effect. The attention weights and modified updated gated are computed as follows, respectively.
\begin{equation}
  \gamma_{i} = \frac{exp((\textbf{q}^{(1)})^\top \textbf{m}_i)}{\sum_i exp((\textbf{q}^{(1)})^\top \textbf{m}_i)},
\end{equation}
% Based on it, the hidden state can be updated as:
\begin{equation}
  \textbf{h}_i = \gamma_{i} \circ \widetilde{\textbf{h}}_i + (1-\gamma_{i}) \circ \textbf{h}_{i-1},
\end{equation}
where $\textbf{h}_{i-1}$ represents the previous hidden state and $\widetilde{\textbf{h}}_i$ represents the internal state of a normal GRU. 

After that, the Bi-GRU is advantageous for retaining the positional and ordering information of contexts.
The final hidden state of this layer serve as the contextual vector to refine the representation of the query, i.e., $\textbf{p}^{(1)} = \textbf{h}_{n}$.
Then, the query can be updated by: 
\begin{equation}
  \textbf{q}^{(2)} = \textbf{W}_{s} [\textbf{p}^{(1)} ; \textbf{q}^{(1)} ] + \textbf{b}_{s},
\end{equation}
where $\textbf{W}_{s} \in \mathbb{R}^{d \times 2d}$ and $\textbf{b}_s$ are the learnable parameters. 
Finally, we adopt the hybrid embedding to represent the rich preference, denoted as  $\textbf{s}$, i.e., $\textbf{s} = \textbf{q}^{(2)}$.

\subsection{Recommendation}
For each session, we predict the next click for all candidate items by multiplying the corresponding item embedding $\textbf{v}_i$.  
% Given the output $\textbf{s}$,
\begin{equation}
  \hat{\textbf{y}}_i = softmax(\textbf{s}^\top \textbf{v}_i ),
\end{equation} 
where $\hat{\textbf{y}}_i$ % \in \mathbb{R}^{l}
 is the probability of the candidate item $v_i \in \mathcal{V}$ to be the next interacted in the session $s$. 
For each session graph, the loss function $\mathcal{L}$ is defined as the cross-entropy of the prediction and the ground truth:
\begin{equation}
  \mathcal{L}(\hat{\textbf{y}}) = -\sum_{i=1}^l  \textbf{y}_i log(\hat{\textbf{y}}_i) + (1 - \textbf{y}_i)  log  (1 - \hat{\textbf{y}}_i),
\end{equation}
where $\textbf{y}$ denotes the one-hot encoding vector of the ground truth item, and $l$ is the number of all candidate items.

% \section{Experiments and Results} 
\section{Experimental Setups} \label{sec:experiment}

\subsection{Datasets and Preprocessing}
We conduct the experiments on three public benchmark datasets, 
i.e., {Yoochoose}, {Diginetica} and {LastFM} datasets.  The statistics are summarized in Table \ref{tab:dataset}.

\textbf{Yoochoose}\footnote{\url{https://2015.recsyschallenge.com/challenge.html}} is a challenging dataset for RecSys Challenge 2015. It contains a stream of user clicks on an e-commerce website within 6 months. 
Following \cite{DBLP:conf/cikm/LiRCRLM17,DBLP:conf/aaai/WuT0WXT19}, we take the recent fractions 1/64 of training sessions and filter out all sessions of length 1 and items that occur less than 5 times.
For generating training and test sets, sessions of the subsequent day are used for testing. 
\textbf{Diginetica}\footnote{\url{https://cikm2016.cs.iupui.edu/cikm-cup/}} comes from CIKM cup 2016. 
We again follow \cite{DBLP:conf/cikm/LiRCRLM17,DBLP:conf/aaai/WuT0WXT19} and filter out all sessions of length 1 and items that occur less than 5 times. 
Sessions of the subsequent week are used as test datasets.
\textbf{LastFM}\footnote{\url{https://www.dtic.upf.edu/ocelma/MusicRecommendationDataset/lastfm{-}1K.html} } is a music recommendation dataset released by \cite{Bertin-Mahieux2011}. Following \cite{DBLP:conf/sigir/WangRMCMR19}, we select the top 40,000 most popular artists as the item set and filter out sessions that are shorter than 2 and longer than 50 items. The splitting of the dataset is the same as the previous work \cite{DBLP:conf/sigir/WangRMCMR19}.

\begin{table}[t]
  \centering
  \caption{Summary of the three benchmark datasets.}
  \label{tab:dataset}
  \resizebox{\linewidth}{!}{$
  \begin{tabular}{|l|r|r|r|r|r|r|}
    \hline
   \multicolumn{1}{|c}{\textbf{Datasets}} & \multicolumn{1}{|c}{\textbf{all the clicks}} & \multicolumn{1}{|c}{\textbf{train sessions}}  & \multicolumn{1}{|c}{\textbf{valid sessions}}  & \multicolumn{1}{|c}{\textbf{test sessions}}  &  \multicolumn{1}{|c}{\textbf{all the items}} & \multicolumn{1}{|c|}{\textbf{avg. length}} \\ 
    \hline
    %  719,470
    {Yoochoose}   &  557,248 & 332,873 & 36,986 &  55,898 & 16,766 & 6.16 \\ \hline 
    {Diginetica}   &  982,961 & 647,532 & 71,947 & 60,858 &  43,097 & 5.12  \\ \hline
    {LastFM}  & 3,804,922 & 26,984    &  5,996   & 5,771  & 39,163    & 13.52 \\ \hline
\end{tabular}
 $}
\end{table}

\subsection{Evaluation Metrics}
We evaluate the recommender system with two commonly-used metrics, {P@20} and {MRR@20}. 
\textbf{P@20} (Precision calculated over top-20 items) computes the proportion of correctly recommended items amongst the top-20 items in an unranking list.
\textbf{MRR@20} (Mean Reciprocal Rank calculated over top-20 items) is the average of reciprocal ranks of the desired items. The reciprocal rank is set to 0 when the rank exceeds 20.

\subsection{Comparison Methods}
We compare our model \textbf{PEN4Rec} with the following baselines:
\textbf{POP} and \textbf{S-POP} always recommend the most popular items in the whole training set or the current session, respectively.
\textbf{BPR-MF} \cite{DBLP:conf/uai/RendleFGS09} uses matrix factorization for recommendation. 
\textbf{FPMC} \cite{DBLP:conf/www/RendleFS10} is a hybrid model for next-basket recommendation. To adapt it to SBRS, we ignore user latent representations when computing recommendation scores.
\textbf{Item-KNN} \cite{DBLP:conf/www/SarwarKKR01} recommends items similar to the existing items in the session, where similarity is based on the co-occurrence number of two items.
\textbf{GRU4Rec} \cite{DBLP:journals/corr/HidasiKBT15} uses RNN with GRUs and session-parallel mini-batch training process.
\textbf{NARM} \cite{DBLP:conf/cikm/LiRCRLM17} further improves GRU4Rec with a neural attention mechanism to capture users' main intent and sequential behaviors.
\textbf{STAMP} \cite{DBLP:conf/kdd/LiuZMZ18} uses the attention mechanism to capture general preference and the recent focus.
\textbf{CSRM} \cite{DBLP:conf/sigir/WangRMCMR19} applies key-value memory networks to consider information from the current session and neighbor sessions.
\textbf{SR-GNN} \cite{DBLP:conf/aaai/WuT0WXT19} encodes the session graph with GGNN and uses an attention layer to represent preferences. 
\textbf{GC-SAN} \cite{DBLP:conf/ijcai/XuZLSXZFZ19} applies self-attention layers after GGNN to capture long-range dependencies.

\subsection{Implementation Details}
Following \cite{DBLP:conf/aaai/WuT0WXT19}, we set the number of layers for GGNN to 1, the dimension of latent vectors to 100.
Besides, we select other hyper-parameters on a validation set.
All parameters are initialized using a Gaussian distribution with a mean of 0 and a standard deviation of 0.1.
The model is trained with the mini-batch Adam optimizer.
We set the batch size to 100, the dropout to 0.5, and the $L_2$ penalty to $10^{-6}$.
The hyper-parameter $k$ is set with range $[1, 5]$ and the best settings are 3, 2, and 4 for Yoochoose, Diginetica, and LastFM, respectively.

\section{Experimental Results and Analysis} \label{sec:results}

\subsection{Results and Analysis}

The general results are presented in Table \ref{tab:result}.
We \textbf{bold} the best performance and \underline{underline} the state-of-the-art result of baselines.
The scores on Diginetica dataset differ from results reported in \cite{DBLP:conf/cikm/LiRCRLM17,DBLP:conf/kdd/LiuZMZ18} because they did not sort the session items according to ``timeframe" field, which ignores the sequential information.

\begin{table}[t]
  \centering 
  \caption{Experimental results (\%) on three datasets.  }
  \resizebox{0.71\columnwidth}{!}{$
  \begin{tabular}{|l|c|c|c|c|c|c|}
    \hline 
    \multirow{2}{*}{\textbf{Methods}}  & \multicolumn{2}{c|}{\textbf{Yoochoose}}  & \multicolumn{2}{c|}{\textbf{Diginetica}}    &  \multicolumn{2}{c|}{\textbf{LastFM}}      \\
    \cline{2-7}
     &  {\textbf{\textit{P@20}}}  & \textbf{\textit{MRR@20}}  
     &  \textbf{\textit{P@20}}  & \textbf{\textit{MRR@20}}  
     &  \textbf{\textit{P@20}}  & \textbf{\textit{MRR@20}}  \\
    \hline
    POP &  6.71 &1.65 & 0.89 &0.20 & 4.43 & 1.15 \\
    S-POP &  30.44 &18.35 & 21.06 & 13.68 & 22.38 & 8.73  \\
    BPR-MF 
    \cite{DBLP:conf/uai/RendleFGS09}  
    &31.31 &12.08& 5.24 &1.98 & 13.38 & 5.73  \\ 
    FPMC 
    \cite{DBLP:conf/www/RendleFS10}
     & 45.62  & 15.01 &26.53& 6.95 & 24.08 & 8.23  \\ 
    Item-KNN 
    \cite{DBLP:conf/www/SarwarKKR01} 
    &  51.60 &21.81 & 35.75 &11.57 & 11.59 & 4.19 \\
    \hline
    GRU4Rec 
    \cite{DBLP:journals/corr/HidasiKBT15} 
    & 60.64 &22.89 & 29.45 &8.33 & 21.42 & 8.21 \\ 
    NARM 
    \cite{DBLP:conf/cikm/LiRCRLM17} 
    & 68.32 &28.63 & 49.70 & 16.17 & 25.64 & 9.18 \\
    STAMP 
    \cite{DBLP:conf/kdd/LiuZMZ18} 
    & 68.74 &29.67 & 45.64 & 14.32 & - & - \\
    CSRM 
    \cite{DBLP:conf/sigir/WangRMCMR19}
     & 69.85 & 29.71 & {51.69} & 16.92 & \underline{27.55} & 9.71 \\
    SR-GNN   
    \cite{DBLP:conf/aaai/WuT0WXT19}
     & {70.57} & \underline{30.94} & {50.73} & {17.59} & 26.20 & {10.43}  \\ 
    GC-SAN \cite{DBLP:conf/ijcai/XuZLSXZFZ19} & \underline{70.66}    &     30.04  &       \underline{51.70}    &   \underline{17.61}     &   26.61       &  \underline{10.62} \\
    \hline   
    \textbf{PEN4Rec}   &\textbf{71.53} & \textbf{31.71} & \textbf{52.50}   & \textbf{18.56} & \textbf{28.82}  & \textbf{11.33} \\
    \textbf{Improve}   &\textbf{1.2\%} & \textbf{2.5\%} & \textbf{1.5\%}   & \textbf{5.4\%} & \textbf{4.6\%}  & \textbf{6.7\%} \\
\hline
\end{tabular}
 $}    
  \label{tab:result}
\end{table}

From the table, we have the following observations: 

1) \textbf{PEN4Rec} obtains the best performance on three datasets, which demonstrates the effectiveness of the proposed model. 
This mainly contributes to the embedding of the session graph and the two-stage modeling of preference evolving process.

2) All deep learning based methods in the second block that make full use of user-item interactions to represent user preferences, are superior to conventional methods in the first block that cannot effectively use the time order.

3) Graph-based models, \textbf{PEN4Rec}, \textbf{SR-GNN}, and \textbf{GC-SAN} consistently outperform most RNN-based models like \textbf{GRURec} and \textbf{NARM}, and the attention-based model \textbf{STAMP}. This proves the graph structure is more suitable for SBRS than the sequence structure, the RNN modeling, or a set structure, the attention modeling.

4) Although RNN-based \textbf{CSRM} obtains worse results than \textbf{SR-GNN} and \textbf{GC-SAN} on most datasets, \textbf{CSRM} outperforms them under the P@20 metric on LastFM dataset. This may be due to the supplement of collaborative neighborhood information.

\subsection{Ablation Study}
In this part, we first compare with the following variants to analyze the key components of \textbf{PEN4Rec}.
\begin{itemize}
\item 
  \textbf{GNN-Last} and \textbf{AGNN-Last} remove the two-stage modeling and recommend based on the embedding of the last item. They use GGNN and attention-based GGNN to encode the session graph, respectively.
  \item
  \textbf{PEN4Rec-Non} removes the second-stage process and recommends based on the output of the first-stage process. 
  \end{itemize}

Besides, we compare with some two-stage variants to show the superiority of the proposed model. 
\begin{itemize}
  \item \textbf{PEN4Rec-ATT} replaces the second-stage process with the attention mechanism where the query is the output of the first-stage process.
  \item
  \textbf{PEN4Rec-GRU} replaces the second-stage process with GRU.
%   3)
    \item 
  \textbf{PEN4Rec-ATT-GRU} applies GRU with attentional input to reason preference instead of updated vectors in the second-stage process. 
\end{itemize}
  
\begin{figure}[t]
  \centering
  \includegraphics[width=\textwidth]{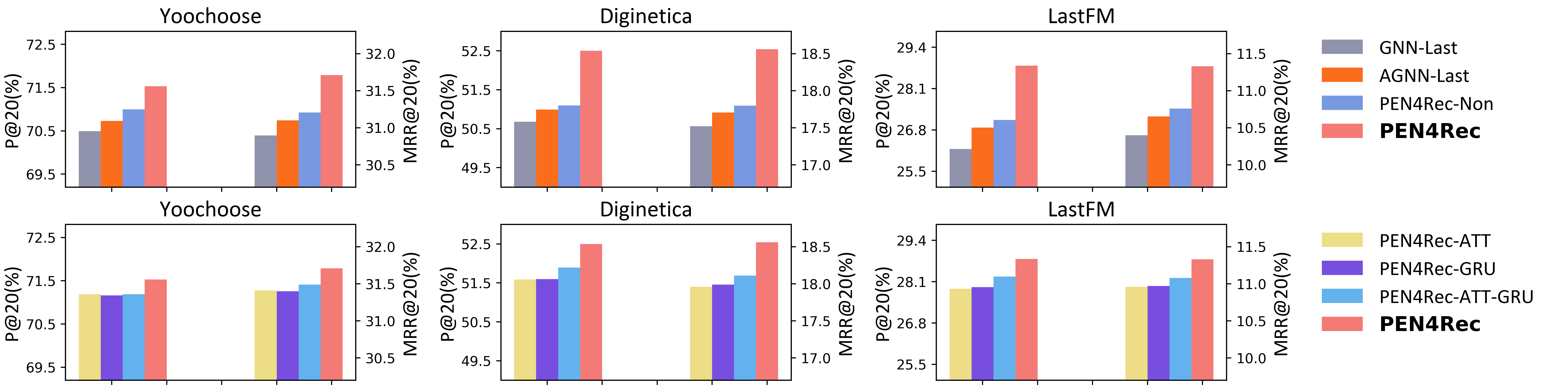} %_hop
  \caption{Results (\%) of ablation study on three datasets.} 
  \label{fig:agg}
\end{figure}
As shown in the top three subgraphs of Fig.~\ref{fig:agg}, \textbf{PEN4Rec} obtains the best performance. \textbf{PEN4Rec-Non} is more powerful than \textbf{AGNN-Last} and \textbf{GNN-Last} which ignore the special modeling of user preferences. The results reveal the effectiveness of the first-stage process.
However, \textbf{PEN4Rec-Non} removes the second-stage retrieval and cannot fully explore latent preferences behind explicit behaviors.  
Its inferior performance than the full model proves the effectiveness of the proposed second-stage process.
Moreover, \textbf{AGNN-Last} outperforms \textbf{GNN-Last}, which proves the effectiveness of the introduction of multi-hop neighbors.
 
As shown in the down three subgraphs of Fig.~\ref{fig:agg}, compared with all two-stage variants, \textbf{PEN4Rec} obtains better results. 
\textbf{PEN4Rec-GRU} deals with dependencies between adjacent behaviors successively and equally, which are not suitable for capturing diverse preferences since the preference has its own evolving track. The method would be disturbed by the preference drifting. 
While \textbf{PEN4Rec-ATT} ignores the sequential features in preferences and suffers from the understanding of evolving preferences.
Although \textbf{PEN4Rec-ATT-GRU} activates relative preferences during preference evolution by the attention score, the variant ignores sequential patterns in different preferences and obtains sub-optimal performance.
Because even zero input can also change the hidden state of GRU, so the less relative preferences also affect the learning of preference evolving.
Different from these variants, we design the second-stage process to combine the local activation ability of the attention mechanism and sequential learning ability from GRU seamlessly. 
In this way, \textbf{PEN4Rec} effectively strengthens the effect of relative sequential behaviors and weakens the disturbance from preference drifting, which boosts modeling the preference evolution. 

\subsection{Parameter Analysis}
We analyze the effect of the vital hyper-parameter $k$ in PEN4Rec. It determines how many recent items are used to form the query that retrieves the relevant behaviors from historical contexts.

\begin{figure}[t]
  \centering
  \includegraphics[width=0.99\columnwidth]{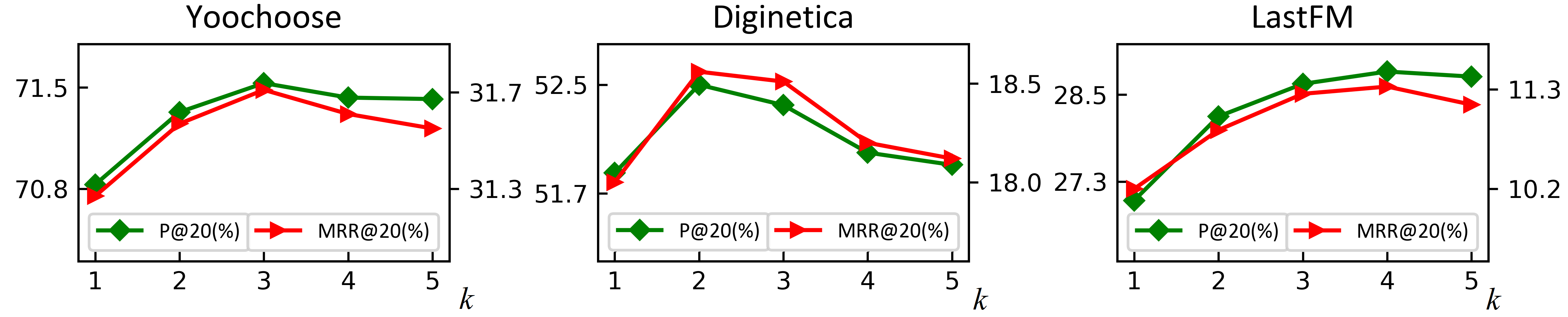}
  \caption{Results against the hyper-parameter $k$. The X-axis indicates the hyper-parameter $k$. The left Y-axis refers to P@20 (\%) and the right Y-axis refers to MRR@20 (\%). The green line represents results in terms of P@20 and the red line denotes results in terms of MRR@20.}
  \label{fig:hop}
\end{figure}

Fig.~\ref{fig:hop} shows results against the hyper-parameter $k$ on three datasets.
We observe that the best settings are 3, 2, and 4 on Yoochoose, Diginetica, and PHEME, respectively.  
This demonstrates that adequate history information indeed contributes to retrieving rich contextual features and providing more clues to reason the preference evolving trajectory. Thereby, more expressive preferences can be captured for more precise recommendations.

\section{Conclusion} \label{sec:conclu}
In this paper, we propose a novel PEN4Rec to model preference evolving process by a well-designed two-stage retrieval from historical contexts for session-based recommendation. Specifically,  the first-stage process integrates relevant behaviors according to recent items. Then, the second-stage process retrieves diverse preferences and models the preference evolving trajectory over time dynamically. The process effectively strengthens the effect of relevant sequential behaviors during the preference evolution and overcomes the disturbance from preference drifting. Experimental results on three benchmark datasets demonstrate the effectiveness and superiority of the proposed model.

% \bibliography{session}

\end{document}